\documentclass[twocolumn,prb,preprintnumbers,amsmath,amssymb]{revtex4}

\usepackage{graphicx,epsf}
\usepackage{dcolumn}
\usepackage{bm}

\begin{document}

\def\eq#1{{Eq.~(\ref{eq:#1})}}
\def\fig#1{{Fig.~\ref{fig:#1}}}
\def\tab#1{{Table~\ref{tab:#1}}}
\newenvironment{bulletList}{\begin{list}{$\bullet$}{}}{\end{list}}

\title{
Computational study of the dielectric properties of [La,Sc]$_2$O$_3$ solid solutions
}
\author{
Hiroyoshi Momida$^{1}$, Eric Cockayne$^{2}$, Naoto Umezawa$^{3}$, and Takahisa Ohno$^{1}$
}
\affiliation{
$^1$Computational Materials Science Center, National Institute for Materials Science, Sengen 1-2-1, Tsukuba, Ibaraki 305-0047, Japan\\
$^2$Ceramics Division, Materials Science and Engineering Laboratory, National Institute of Standards and Technology, Gaithersburg, MD 20899-8520 USA\\
$^3$Advanced Electronic Materials Center, National Institute for Materials Science, Namiki 1-1, Tsukuba, Ibaraki 305-0044, Japan
}

\date{February 5, 2010}

\begin{abstract}
First-principles calculations were used to compute the
dielectric permittivities of hypothetical [La,Sc]$_2$O$_3$ solid solutions
in the cubic (bixbyite) and hexagonal La$_2$O$_3$ phases.
Dielectric enhancement is predicted at small Sc concentrations due to
the rattling ion effect. Similar calculations for a model amorphous
La$_2$O$_3$ structure show little change in permittivity when a small
amount of Sc is substituted for La. In this case, the
local environment around the Sc changes in a way that compensates
for the rattling ion effect.
\end{abstract}

\maketitle

High permittivity (``high $\kappa$") dielectric materials for microelectronics
applications require a combination of properties,
including a permittivity $\kappa$ larger than that of amorphous SiO$_2$
($\kappa$ = 3.9), thermodynamic stability in contact with Si,
and sufficient band offsets with respect to Si. \cite{Wilk01}
While high-$\kappa$ dielectrics based on HfO$_2$ have been incorporated
into commercial computer chips, \cite{Matthews08}
other ``higher-$\kappa$" materials continue to be actively
investigated, \cite{Robertson06} as future Metal-Oxide-Semiconductor
Field-Effect Transistor (MOSFET) scaling will likely require
dielectrics with higher $\kappa$ than HfO$_2$. \cite{Robertson08}
An additional concern for dielectrics for MOSFET applications is
whether the dielectric is amorphous. Amorphous dielectrics are considered
superior, as microcrystalline dielectrics can exhibit significant
charge leakage and ionic diffusion due to grain boundaries. \cite{Wilk01}
On the other hand, crystalline dielectrics are still of interest,
as epitaxial dielectric growth \cite{Robertson06} would avoid the above problems.

Early compilations of candidate high-$\kappa$ materials indicated that
La$_2$O$_3$ has higher permittivity than HfO$_2$, \cite{Wilk01} thus much
attention on higher-$\kappa$ materials has focused on La$_2$O$_3$ and other
$R_2$O$_3$-based materials ($R$ = rare earth). High permittivities have
been reported in both single component $R_2$O$_3$ compounds, such as
Sm$_2$O$_3$, \cite{Yang08} and in stoichiometric $R$ScO$_3$ compounds, such as
GdScO$_3$ and DyScO$_3$. \cite{Zhao05} This work explores a possible way to further enhance
the permittivity of rare-earth dielectrics: nonstoichiometric substitution of
Sc for the $R$ ion. This idea is inspired by recent theoretical work on
the Ba$_{1-x}$Ca$_x$ZrO$_3$ (BCZO) perovskite system. \cite{Bennett08}
BCZO has a dielectric anomaly, where the permittivity is higher for
intermediate compositions than for either endmember, reaching a
maximum in the range $x \approx$ 0.1 to 0.2. \cite{Yamaguchi80,Levin03}
The enhancement in permittivity is due to a ``rattling ion" effect:
the smaller Ca ion in the lattice site of the larger Ba ion is relatively loosely bound
and can thus move farther under an applied electric field.
A similar effect might be expected when a small Sc ion substitutes
isovalently for a larger $R$ ion in $R_2$O$_3$ compounds. The largest $R^{3+}$ ion is
La$^{3+}$ (Ref.~\onlinecite{Gmelin74}); thus we explore the dielectric properties of
the La$_{2-2x}$Sc$_{2x}$O$_3$ (LSO) solid solution system in this work.

La$_2$O$_3$ has two room-temperature polymorphs: a cubic ($c$)
bixbyite phase and a hexagonal ($h$) phase. \cite{Wyckoff64,Gmelin74}
Due to the small difference in energy between $c$-La$_2$O$_3$ and
$h$-La$_2$O$_3$ (which is thermodynamically stable at sintering temperatures of order
1000 $^{\circ}$C), it is not clear which phase is the most
stable at room temperature. \cite{Gmelin74} We thus explore both
structures. Sc$_2$O$_3$ has the cubic bixbyite structure at room temperature.
LSO has one ordered intermediate phase, \cite{Gmelin74}
with the perovskite ($p$) structure. \cite{Trzebi65}

We ran first principles calculations on the 40-atom primitive cell for
$c$-La$_2$O$_3$ cell, and on a 40 atom $h$-La$_2$O$_3$ supercell
(each cell parameter doubled), for both pure
La$_2$O$_3$ and for the same cells with one Sc substituted for La ($x=0.0625$).
The initial atomic positions were taken from the literature. \cite{Wyckoff64}
Additional calculations were performed for:
(1) $c$-LSO with $x=0.25$, to test the linearity of $\kappa$ vs. $x$;
(2) $c$-LSO with $x=0.9375$ (``reverse" Sc $\rightarrow$ La substitution);
(3) ideal $p$-LaScO$_3$;
(4) a model for amorphous La$_2$O$_3$, and
(5) a model for amorphous LSO ($a$-LSO) with $x=0.0625$.
All calculations were performed at the experimental volume or the
volume given by Vegard's law: $V = (1-x) V_{\rm{La_2O_3}} + x\,V_{\rm{Sc_2O_3}}$.

\begin{figure}[b]
\includegraphics[width=85mm]{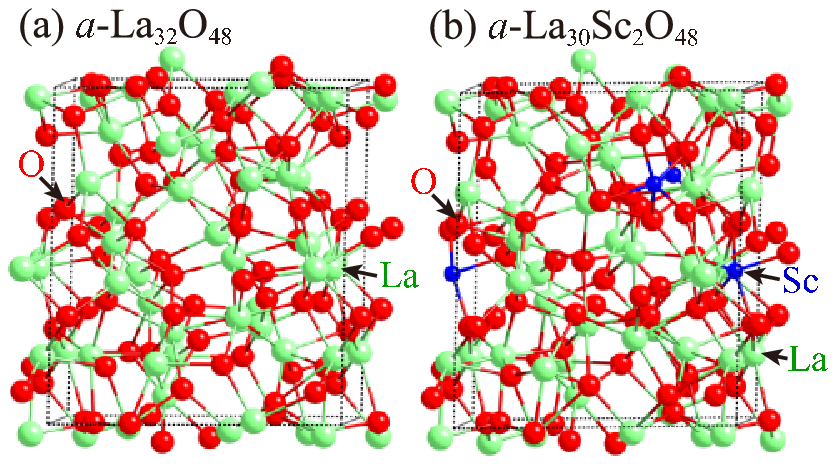}
\caption{(Color online) Relaxed models for (a) amorphous La$_2$O$_3$
and (b) La$_{2-2x}$Sc$_{2x}$O$_3$ ($x=0.0625$).}
\label{fig:amorph}
\end{figure}

Structural relaxations and dielectric constant calculations were performed using
the density functional theory code \textsc{PHASE}. \cite{PHASE}
Ultrasoft pseudopotentials \cite{Vanderbilt90} were used, with the PBE generalized gradient
approximation \cite{Perdew97} for the exchange-correlation functional,
and a plane wave basis set with a 36 Rydberg cutoff energy for the
electronic wavefunctions.
Brillouin zone integrations were performed with the $\Gamma$ point only
for the force calculations, and were done with appropriate grids for
the electronic dielectric constant and the Berry phase calculations.
All structures were relaxed until the force on each atom was smaller
than 0.002 Ry/Bohr.
For $c$-La$_2$O$_3$, there are two inequivalent La positions: $8(b)$ and $24(d)$
(Ref.~\onlinecite{Wyckoff64}; modern convention for space group $Ia\overline{3}$
Wyckoff positions from Ref.~\onlinecite{Inter02}).
At both $x=0.0625$ and $x=0.9375$, the lowest energy is found when the minority
species is on the $8(b)$ site.
Because a single occupied $8(b)$ is lower in energy than $24(d)$, it seems reasonable to assume that, for lowest energy configurations, the $8(b)$ sites fill with Sc before the $24(d)$ sites.
For the $x=0.25$ calculation, therefore it was assumed that lowest energy occurs when the
$8(b)$ site is fully occupied.

\begin{figure}
\includegraphics[width=55mm]{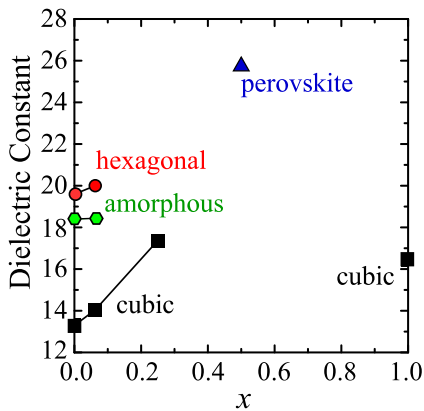}
\caption{(Color online) Comparative computed dielectric constant for
La$_{2-2x}$Sc$_{2x}$O$_3$ structures studied in this work.}
\label{fig:kappa}
\end{figure}

\begin{table}
\caption{Calculated dielectric constant ($\kappa={\kappa}^{\rm{el}}+{\kappa}^{\rm{ion}}$)
and band gap at $\Gamma$ ($E_{\rm{g}}$) for La$_{2-2x}$Sc$_{2x}$O$_{3}$ structures
studied in this work.}
\begin{ruledtabular}
\begin{tabular}{ccccc}
 $x$ & $\kappa$ & ${\kappa}^{\rm{ion}}$ & ${\kappa}^{\rm{el}}$ & $E_{\rm{g}}$ (eV) \\ \hline
 Cubic bixbyite & & & & \\
 0 & 13.29 & 9.13 & 4.17 & 3.67 \\
 0.0625 & 14.04 & 9.83 & 4.22 & 3.36 \\
 0.25 & 17.37 & 13.04 & 4.33 & 2.55 \\
 0.9375 & -- & -- & 4.19 & 3.63 \\
 1 & 16.46 & 12.33 & 4.13 & 3.87 \\
 Hexagonal & & & & \\
 0 & 19.60 & 14.88 & 4.72 & 3.73 \\
 0.0625 & 20.00 & 15.31 & 4.69 & 3.56 \\
 Amorphous & & & & \\
 0 & 18.42 & 14.00 & 4.42 & 2.90 \\
 0.0625 & 18.43 & 14.03 & 4.39 & 2.96 \\
 Perovskite & & & & \\
 0.5 & 25.68 & 20.96 & 4.72 & 3.91 \\
\end{tabular}
\end{ruledtabular}
\label{tab:kappatab}
\end{table}

A model for the La$_2$O$_3$ amorphous phase was generated
by the {\it ab initio} molecular dynamics
``melt and quench" technique. \cite{XZhao05}
A 80-atom cubic La$_2$O$_3$ cell was ``melted" at 3000 K until the
root mean square displacement of oxygens from their original
positions was greater than the original oxygen-oxygen near-neighbor
distance. Then the cell was quenched for 1 ps of simulation time at
1000 K. The lowest energy structure during this run was recorded,
and subsequently a full relaxation (cell shape and atomic positions)
was performed to obtain an $a$-La$_2$O$_3$ model.
(The melt-and-quench technique was done using the \textsc{VASP}
package \cite{Kresse96} with projected augmented wave functions \cite{Blochl94}
and the local density approximation for exchange-correlation; the final
relaxation and subsequent steps used \textsc{PHASE}, as above.)
After this relaxation, bond valence sums \cite{Brown85,Brese91}
were computed to identify the most underbonded La as the
most favorable positions for Sc.
The Sc was placed on the most underbonded La site
because the smaller ionic radius of Sc favors
the lower coordination number.
The two most underbonded
La separated by more than $a/2$ were replaced by Sc, and the
structure then fully relaxed under the new cell volume to
obtain a model for $a$-LSO; $x=0.0625$. The procedure insures that
the $a$-LSO model is as similar as possible to $a$-La$_2$O$_3$ model
for direct comparison of their properties (\fig{amorph}).

The dielectric properties were computed using the frozen phonon
method, as discussed in Ref.~\onlinecite{Cockayne00}. Briefly,
the static dielectric constant of a crystal is given by
\begin{equation}
\kappa = \kappa^{\rm{el}} + \kappa^{\rm{ion}}
= \kappa^{\rm{el}} + C \sum_{\mu} \frac{{Z_{\mu}}^2}{{\omega_{\mu}}^2},
\label{eq:kappa}
\end{equation}
where $C$ is a constant, $\kappa^{\rm{el}}$ the electronic contribution
to the dielectric constant, $\kappa^{\rm{ion}}$ the ionic contribution,
$Z_{\mu}$ the dipole moment associated with mode $\mu$ and $\omega_{\mu}$
its frequency. Aside from obtaining the dielectric constants, the breakdown
of the lattice dielectric contribution over individual modes allows one
to find the origins of any dielectric anomaly that is due to lattice effects.
A rattling ion effect will show up as a significant contribution to $\kappa$
from a mode with low frequency and/or high $Z_{\mu}$ that is located on
or near the rattling ion.

\begin{figure}
\includegraphics[width=85mm]{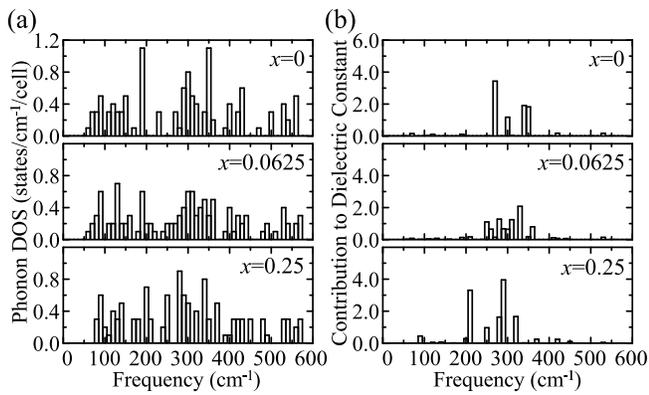}
\caption{(a) Comparative phonon density of states for $c$-La$_{2-2x}$Sc$_{2x}$O$_3$
versus $x$. (b) Contribution of phonons to dielectric constant, versus $x$,
binned by 10 cm$^{-1}$ range of frequency.}
\label{fig:pdos}
\end{figure}

\fig{kappa} shows the dielectric constant computed for each
structure. \tab{kappatab} shows, in addition, $\kappa^{\rm{el}}$, $\kappa^{\rm{ion}}$, and
the calculated bandgaps. For small $x$, the dielectric constants of both
$c$-LSO and $h$-LSO are higher than the weighted average of the
endmember La$_2$O$_3$ and Sc$_2$O$_3$ compounds.
Phonon density of states calculations (\fig{pdos}) show that,
as $x$ increases, the frequencies of the modes around
200 cm$^{-1}$ to 300 cm$^{-1}$ that contribute significantly to
the dielectric constant decrease. The dielectric constant is thus
enhanced, per \eq{kappa}. \cite{foot1}
Analysis of these lower-frequency phonons shows that the eigenvectors
are dominated by Sc motion coupled to motion of its nearest neighbor O octahedron
in the opposite direction. We therefore conclude that the
enhancement of $\kappa$ is due to the rattling ion effect.
This hypothetical effect is distinct from the enhancement of
permittivity due to rare-earth doping of HfO$_2$, which is believed
to result from the stabilization of the HfO$_2$ tetragonal phase relative
to the monoclinic one. \cite{Govi07}
For the $h$-LSO case, the substituted Sc atom displaces along the hexagonal
$c$ axis to optimize the structure. Tight (0.201 nm) and weak (0.304 nm)
Sc-O bonds are formed. As a result, Sc moves more easily along the
$c$ axis, but is more rigid in the $ab$ plane. The $\kappa_{cc}$ component
is enhanced, while $\kappa_{aa}$ and $\kappa_{bb}$ are reduced. Overall,
$\kappa$ is enhanced.
When a small amount of La is substituted for Sc in $c$-Sc$_2$O$_3$, the
$\kappa^{\rm{el}}$ increases, which we attribute to the bandgap reduction
(\tab{kappatab}), but $\kappa^{\rm{ion}}$ was not calculated, as a mode
with imaginary frequency ({\it i.e.} an instability) appears.
In other words, a small amount of La substitution is predicted to
break the cubic symmetry.
The relative increase in $\kappa$ with composition is less than that
found for BCZO, \cite{Bennett08} and the dielectric
constant predicted at $x=0.25$ does not exceed that computed for
the intermediate perovskite LaScO$_3$ phase. Nonetheless, if a phase
that is close in composition to $R_2$O$_3$ is desired, our results
suggest a way to increase the permittivity with relatively small
doping of Sc. Furthermore, if it were possible to substitute a small
amount of Sc on the $A$ site of a $R$ScO$_3$ perovskite (an isostructural
analog to BCZO) without destabilizing the perovskite phase, the rattling ion
effect might yield a permittivity even higher than for
stoichiometric $p$-$R$ScO$_3$ compounds.

Our results predict that $\kappa$ for amorphous La$_2$O$_3$
is higher than for $c$-La$_2$O$_3$ but smaller than for
$h$-La$_2$O$_3$. Delugas {\it et al.} predict a similar permittivity
enhancement for amorphous Sc$_2$O$_3$ with respect to
$c$-Sc$_2$O$_3$. \cite{Delugas08}
In contrast to the crystalline cases studied, the computed dielectric constant of
$a$-La$_2$O$_3$ does not increase when a small amount of
Sc is substituted for La.
We explain the lack of any rattling ion effect as a consequence
of different Sc and La local environments in $a$-LSO.
Upon relaxation, the Sc ions have on average 5.0 oxygen neighbors
at an average distance of 0.208 nm, while the La ions have on average
6.0 oxygen neighbors at an average distance of 0.246 nm.
The results suggest that the local environment around Sc in $a$-LSO
``tightens up" relative to the typical La local environment in a way
that compensates for the rattling ion effect.

To summarize, small substitution of Sc for La in crystalline
La$_2$O$_3$ compounds is predicted to cause an anomalous increase
in the dielectric constant due to the rattling ion effect: a
small Sc ion in a loose environment can move farther in an applied
electric field. For amorphous La$_2$O$_3$, we predict
no anomalous increase, as La and Sc have different local environments
in this case.

We thank Toyohiro Chikyow for useful discussions.
N. U. thanks P. Weakliem for technical advice on using computer
facilities in the California Nanosystems Institute (CNSI) at
University of California Santa Barbara. Some of the calculations
were performed on Hewlett-Packard clusters in CNSI.

\end{document}